\documentclass[journal=jacsat,manuscript=article]{achemso}

\usepackage[T1]{fontenc} 

\author{Koushik R. Das}
\author{Sudipta Dutta}
\email{sdutta@labs.iisertirupati.ac.in}
\phone{+91 877 2500 434}
\affiliation{Department of Physics, Indian Institute of Science Education and Research (IISER) Tirupati, Tirupati-517507, Andhra Pradesh, India}

\title[An \textsf{achemso} demo]
  {Asymmetric Electronic Transport in Porphine: Role of Atomically Precise Tip-Electrode}

\abbreviations{IR,NMR,UV}

\begin{document}

\begin{abstract}
   Electronic conductance through a single molecule is sensitive towards its structural orientation between two electrodes, owing to the distribution of molecular orbitals and their coupling to the electrode levels, that are governed by quantum confinement effects. Here, we vary the contact geometry of the porphine molecule by attaching two $Au$ tip electrodes that resemble the mechanical break junction, via thiol anchoring groups. We investigate the current-voltage characteristics of all the contact geometries using non-equilibrium Green's function formalism along with density functional theory and tight-binding framework. We observe varying current responses with changing contact sites, originating from varied wave-function delocalization and quantum interference effect. Our calculations show asymmetric current-voltage characteristics under forward and reverse biases due to structural asymmetry of the tip electrodes in either sides of the molecule. We establish this phenomenon as a universal feature for any molecular electronic device, irrespective of the inherent structural symmetry of a molecule. This will provide fundamental insights of electronic transport through single molecule in real experimental setup. Furthermore, our observations of varying current response can further motivate the fabrication of sensor devices with porphine based biomolecules that control important physiological activities, in view of their applications in advanced diagnostics.
\end{abstract}

\section{Introduction}
	Single molecule based electronic devices are regarded as the possible alternative to the conventional semiconductor-based devices, in view of their higher efficiency in sub-nanometer length scale. Such devices can perform in ultra-low energy regime without significant Joule heating. This has garnered widespread attention in the past few decades since the founding works of Aviram and Ratner\cite{Ratner1974}, followed by further theoretical and experimental developments\cite{landauer1992conductance,buttiker1986four,datta1997electronic,datta2000nanoscale}. Overwhelming advances in experimental techniques in recent times, such as, scanning tunneling microscopy (STM) and mechanically-controlled break junction (MCBJ),  have evolved this field from theoretical hypotheses to experimental materiality\cite{homma2023dependence}. These pave new avenues to exploit the quantum states of a molecular system and tailor them to suit targeted requirements. Molecular transistors, diodes, switches are some devices which have already been proposed and experimentally demonstrated based on molecular systems\cite{metzger2018quo,joachim2000electronics}. 

    Acyclic and cyclic $\pi$-conjugated systems are preferred for molecular electronics devices, due to the presence of delocalized $\pi$-electrons. Few alkene and alkyne based systems show electron transport through their acyclic chains\cite{yelin2016conductance,malen2009identifying,baer2002phase,mendez2011surface,crljen2005nonlinear}. Among the cyclic systems, some of the aromatic molecules, such as benzene\cite{choi2005role}, naphthalene\cite{dutta2008comparative}, azulene\cite{kala2018first} etc. and a few non-aromatic systems like isoxasole \cite{langer1987toward} and B$_2$C$_2$N$_2$H$_6$\cite{ghosh2021small} have been reported for electron transport through the molecular backbone.	It has been observed that positioning electrodes at different contact sites of a molecule imparts distinct electronic responses, because of electronic phases \cite{lambert2015basic} in respective molecular orbitals and their interferences. This was shown for several systems, e.g., benzene, naphthalene and different annulene molecules\cite{ke2008,gunasekaran2020,yoshizawa2008orbital}.

    Here, we explore the electron transport through porphine molecule, a building block for many naturally occurring porphyrin biomolecules such as haemoglobin, chlorophyll, Vitamin-B$_{12}$ etc. It is a tetrapyrrole, comprising of four pyrrole rings connected by methine (R-CH=R$^\prime$) bridges. It is an aromatic molecule with rigid planar geometry and high chemical stability that are desirable traits for an efficient molecular device. The $\pi$-conjugated ring structure provides multiple possible electrode contact sites, that is expected to result in varying interferences of molecular orbitals. The proposed applicability for studying such systems is to develop biochemical sensors, since their electronic responses are expected to vary in accordance with changes in their immediate environment, which can have wide applicability in pathogen and toxic detection in futuristic devices\cite{hong2012molecular,fagadar2018sensors,mirkin1992molecular,harashima2022unique}.

    Porphine have been found to coordinate with various transition metal ions (metalloporphyrins), which affects the electron transport through the overall system, as theoretically established by Shuai et.al.\cite{ouyang2007quantum} and Wang et.al.\cite{wang2009theoretical}. The electronic behavior has also been theoretically considered in various geometric forms, such as, tapes\cite{tagami2003electronic}, wires\cite{zhao2010molecular,li2013}, and sheet \cite{xie2020two}. The effect of donor-acceptor groups on the electron transport behavior of porphyrin molecule was investigated earlier, considering two specific molecular orientations\cite{long2008negative,li2011}.

    In this article, we broadly investigate ten possible orientations of the porphine molecule (see Supporting Information (SI)), sandwiched between two gold electrodes, mimicking the experimental situation of  mechanically controlled break junction (MCBJ). We show all the possible electrode contact sites marked with $``a"$ and $``b"$ with numerical values in Figure 1a. We fix $``a_{1}"$ and $``b_{1}"$ on one of the methine bridges and on one of the pyrrole rings, respectively, and vary the position of the second electrode marked with subsequent numerical values. We present the results of the $``a"$-type contact sites in the main manuscript and provide the results of the $``b"$-type contact sites in the SI. In Figure 1b, we present the schematic of the device geometry with the molecule placed in between two semi-infinite bulk gold (Au) electrodes. As can be seen, we consider thiol anchoring group, owing to its high binding affinity with the Au electrode\cite{tsuji2011orbital,delamarche1996golden}. For precise attachment of the desired contact sites of the molecule, one needs the atomic precision of the electrodes, as used in scanning tunneling microscopy (STM) experimental setup. This drives us to consider two pyramidal gold tips consisting of four Au atoms in either sides of the molecule. Note that, these tips are sculptured from the bulk gold electrodes in either sides. This approach allows us to probe larger number of contact geometries as compared to only two specific configurations reported so far, based on the use of bulk electrodes\cite{wang2009theoretical,li2011}. Recent experimental sophistication can provide such atomically precise measurements that can emulate the results reported here.

    We observe diverse current responses under positive and negative biases, that significantly depends on the contact site orientations. To gain insight into such behavior, we analyze the transmission functions and molecular orbitals for each setup. All calculations are done using the non-equilibrium Green's function (NEGF) formalism with density functional theory (DFT) and tight-binding (TB) framework.

\section{Computational Method}

  \begin{figure*}\centering		
	\includegraphics[scale=0.25]{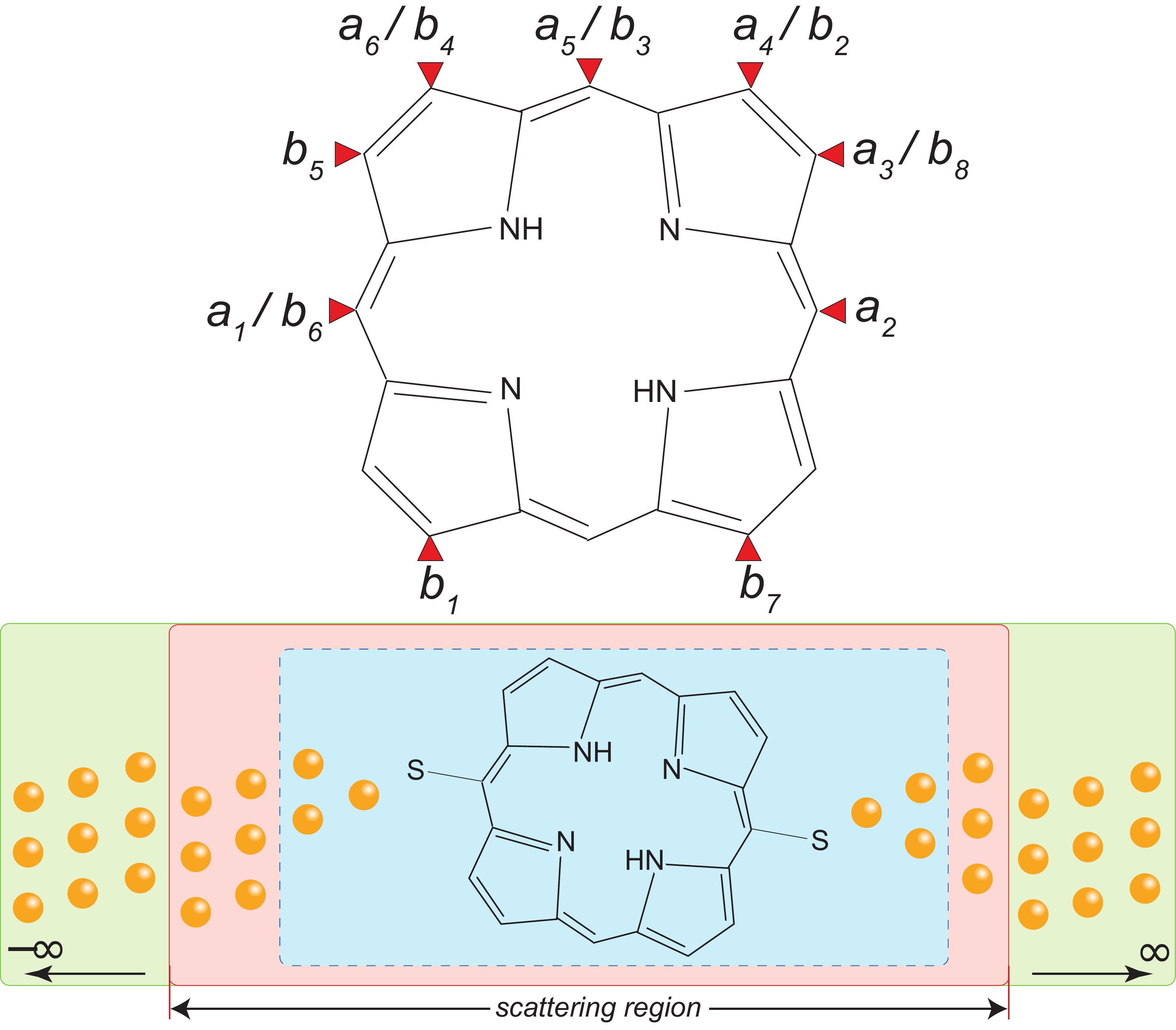}
	\caption{(a) The possible electrode contact sites (red arrow heads) on porphine molecule with both $``a"$ and $``b"$ types of attachments. (b) The computational cell, used for transport calculations of $a_1-a_2$ contact geometry. The molecule is attached to the tip electrodes via thiol anchoring groups in both sides. This is considered as extended molecule and depicted by the blue box with dashed line. The tip electrodes are sculptured out from the bulk $Au$ electrodes. Parts of the bulk electrodes in both sides are considered to be inside the scattering region (red box with solid line) along with the extended molecule. The bulk electrode is modeled with 27 $Au$ atoms, arranged in 3 layers in crystallographic (111) direction, with semi-infinite boundary condition.}
  \end{figure*}

   The geometries of the porphine molecule and the molecule with two thiol anchoring groups at varying contact sites ($``a"$ and $``b"$) are first optimized using double zeta polarized (DZP) basis within DFT framework, as implemented in SIESTA-3.2\cite{soler2002siesta}. We consider Perdew-Burke-Ernzerhoff (PBE) exchange-correlation with generalized gradient approximation (GGA)\cite{perdew1996generalized}. Then we construct the extended molecule by attaching the Au tips to the anchoring groups and relaxed the geometry further with same parameterizations. For transport calculations within  Landauer-B\"{u}ttiker formalism, we adopt the NEGF-DFT approach as implemented in TranSIESTA\cite{stokbro2003transiesta}. The NEGF formulation specifically deals with the non-equilibrium situation that happens under the application of non-zero bias\cite{datta1997electronic}. Each semi-infinite metallic electrode in either sides of the extended molecule is modeled with 27 Au atoms. For all the transport calculations, the computational cell consists of three components, namely, left electrode, extended molecule and right electrode (see Figure 1b). We consider single zeta (SZ) basis for the Au atoms and double zeta (DZ) basis for all the other atoms within PBE-GGA parameterization. It has been shown that the choice of basis sets gives some quantitative difference, without much variation of the qualitative features\cite{cohen2007}. This combination of basis sets is chosen to reduce the computational expense for large size of the systems. The pseudopotentials for all the atoms are generated within Troullier-Martins scheme\cite{troullier1991efficient}. The further details of the computational parameters are provided in the SI.

   The current is calculated by integrating the transmission function, $T(E,V)$, as given by the Landauer-B\"{u}ttiker formula:

  \begin{equation}
    I(V)= {\frac{2e^{2}}{h}} \int_{\mu_{R}}^{\mu_{L}} T(E,V) dE 
  \end{equation}

   \noindent where $e$ is the electronic charge, $h$ is Planck's constant, $V$ is applied bias and $E$ is energy of electron. {\large$\mu$\small$_{L/R}$} are the potentials of the $L/R$ electrodes, defined as \large${\mu}$\small$_{L/R} = E_F \pm eV/2 $, with \large$\mu$\small$_L$ - \large$\mu$\small$_R=eV$; $E_F$ being the Fermi energy of the electrode at zero-bias. The interval $[\mu_R(V),\mu_L(V)]$ i.e $[-V/2,V/2]$ denotes the energy region that contributes to the current integral and denoted as the bias window. This is in accordance with the fact that electrons near the $E_F$ will only contribute to the total current. 

   The $T(E,V)$ is defined as the total probability of an electron with energy $E$ to traverse between electrodes through the device region (see Figure 1b) and can be expressed as:
   \begin{equation}
    T(E,V)=Tr[\Gamma_L(E,V)G(E,V)\Gamma_R(E,V)G^\dagger(E,V)]
  \end{equation} 

  \noindent In this expression, $G(E,V)$ is the retarded Green's function:
  \begin{equation}
    G(E,V)=[(E+i\eta)I-H]^{-1}=[(E+i\eta)I-H_{mol}-\Sigma_L-\Sigma_R]^{-1}
  \end{equation}
  \noindent where $H$ is the device Hamiltonian, comprising of molecular Hamiltonian $H_{mol}$ and self energies $\Sigma_{L/R}$, that arise due to the coupling of the molecule with $L/R$ electrodes. $I$ is the identity matrix and $\eta$ is an infinitesimally small number to avoid numerical divergence. The spectral density, $\Gamma_{L/R}$ that provides the eigenstate broadening due to $L/R$ electrodes can be expressed in terms of the self-energies as, 
  \begin{equation}
    \Gamma_{L/R}=i~ (\Sigma_{L/R}-\Sigma^\dagger_{L/R})
  \end{equation}

  \noindent We vary the external bias from $-1V$ to $+1V$ with an interval of $0.1V$ and calculate the transmission function and the current for each bias.

\section{Results and Discussions}

    The current response of the devices with varying $``a"$-type contact geometries are shown in Figure 2a, as a function of the applied bias in both forward ($+ve$) and reverse ($-ve$) directions. One can see the varying response based on the contact site variations. However, the current shows peak at $0.1V$ for all the contact geometries, followed by a steep decrease. The current peaks again at an elevated external bias. The $a_1-a_2$ contact geometry shows maximum current, whereas, the $a_1-a_4$ exhibits negligible current response. Interestingly, the current responses for all the systems under forward and reverse biases show asymmetric behavior. It is noteworthy that, the previous studies reported a linear current-voltage ($I-V$) characteristics for $a_1-a_2$ under the forward bias\cite{wang2009theoretical,li2011} only. However, such linear response was obtained by placing the molecule in between two semi-infinite bulk electrodes, without considering the tip-leads that we consider in our study. We shall discuss about the implication of that in later part.

    \begin{figure}
    	\includegraphics[scale=0.2]{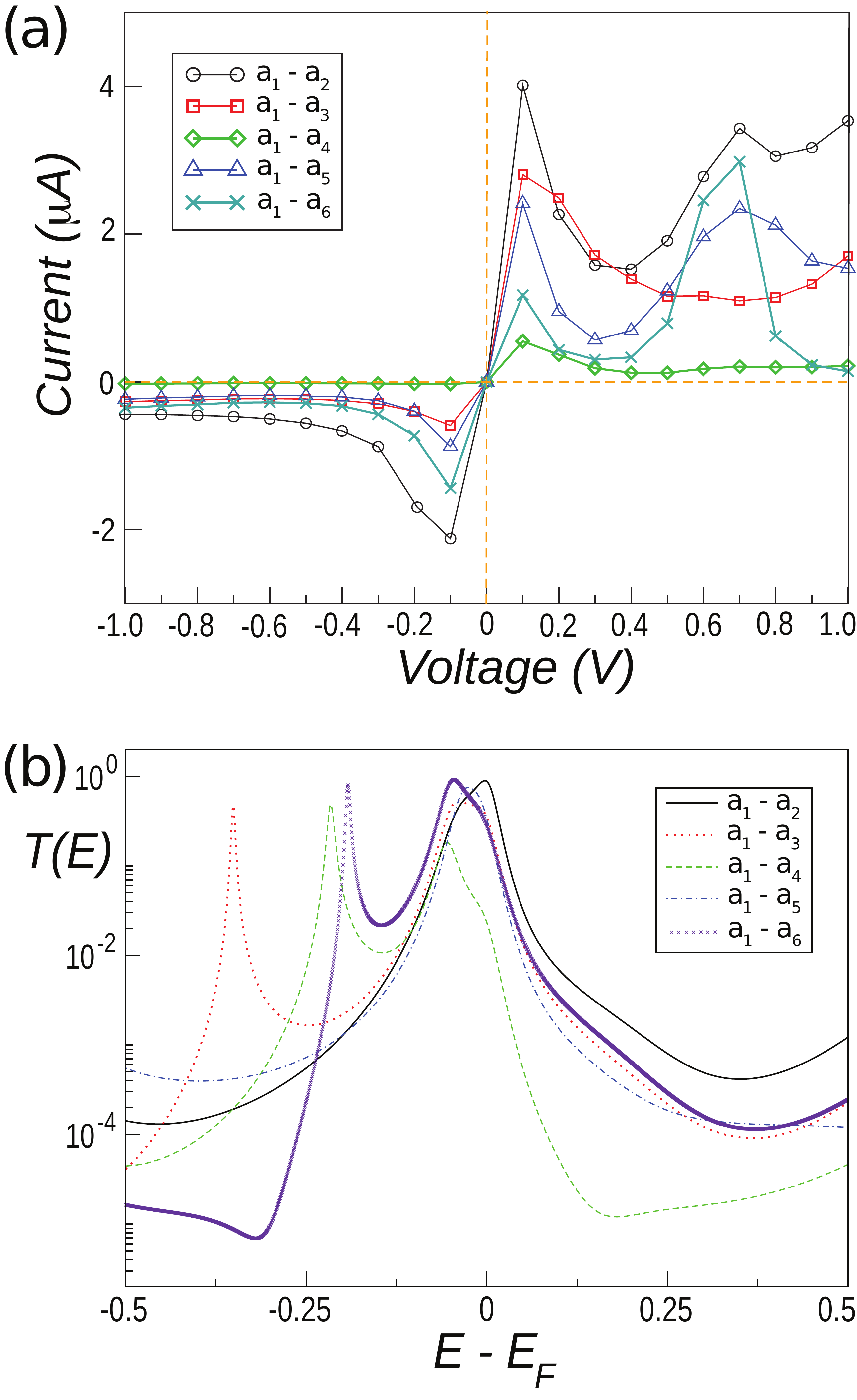}
    	\caption{(a) The current-voltage characteristics of porphine molecule with all possible $``a"$ type contact geometries for both forward ($+ve$) and reverse ($-ve$) biases. (b) The zero bias transmission of those contact geometries. The transmission values are depicted in a normalized logarithmic scale and the energy is scaled with respect to the electrode Fermi energy.}
    \end{figure}

    To gain insight about the transport behavior, we further show the normalized zero bias transmissions for all the $``a"$-type contact geometries in Figure 2b. As can be seen, all the systems show a transmission peak very near to the Fermi energy. This leads to a sharp increase in the current for very small applied bias. That is why, we observe the peak of current at $0.1V$ for all the systems. In case of $a_1-a_4$ system, the transmission peak is little away from the Fermi energy and also peak height is smaller as compared to that of other systems. This indicates the very small current response that we have observed for this system. 

    In order to identify the contribution from the molecular orbitals (MOs) to the transmission, we further calculate the molecular projected self-consistent Hamiltonian (MPSH) eigenstates by diagonalizing the scattering region Hamiltonian\cite{stokbro2003theoretical,li2011}. This effectively projects the total device states on the scattering region, shown in Figure 1b. These calculations are done using the Inelastica package\cite{paulsson2007transmission} without inclusion of the bulk electrodes. The MPSH eigenstates can be termed as molecular conductance orbitals (MCOs). We present the MCOs for $a_1-a_2$ contact geometry as the vertical dotted lines in Figure 3, along with the corresponding normalized zero bias transmission. Note that the MCOs with higher delocalization of wave-function over the whole scattering region contribute largely towards the transmission. That is why, we present only selective plots of MCOs in Figure 3. For comparison, we also present the corresponding molecular orbitals (MOs) for the isolated molecule. Weak electrode-molecule coupling results in similar appearances of the MOs and MCOs. 

    As can be seen, the transmission at the Fermi energy is originated from the highest occupied molecular orbital (HOMO) of the porphine molecule with $a_1-a_2$ contact sites. The other notable transmission peaks arising from the HOMO-2 and LUMO+1 (lowest unoccupied molecular orbital+1) are energetically quite distant from the Fermi energy and unlikely to take part in the electron conduction at small bias values, as considered in our study. This indicates a HOMO dominated electronic transport that is consistent for other contact geometries as well. The MCO corresponding to the HOMO-1 is almost degenerate with that of the HOMO-2. However, it does not contribute to the transmission peak due to the localization of the wave-function only on the molecular region. Similar wave-function localization on the molecular region in the MCO corresponding to the LUMO restricts its contribution towards transmission. Note that, metal-induced gap states result into MCOs (vertical dotted lines) in between two consecutive MOs. Due to substantial wave-function localization on the metal regions only, these MCOs do not contribute towards the overall conduction, as evident from the corresponding small transmission values.

    \begin{figure}
    	\includegraphics[scale=0.3]{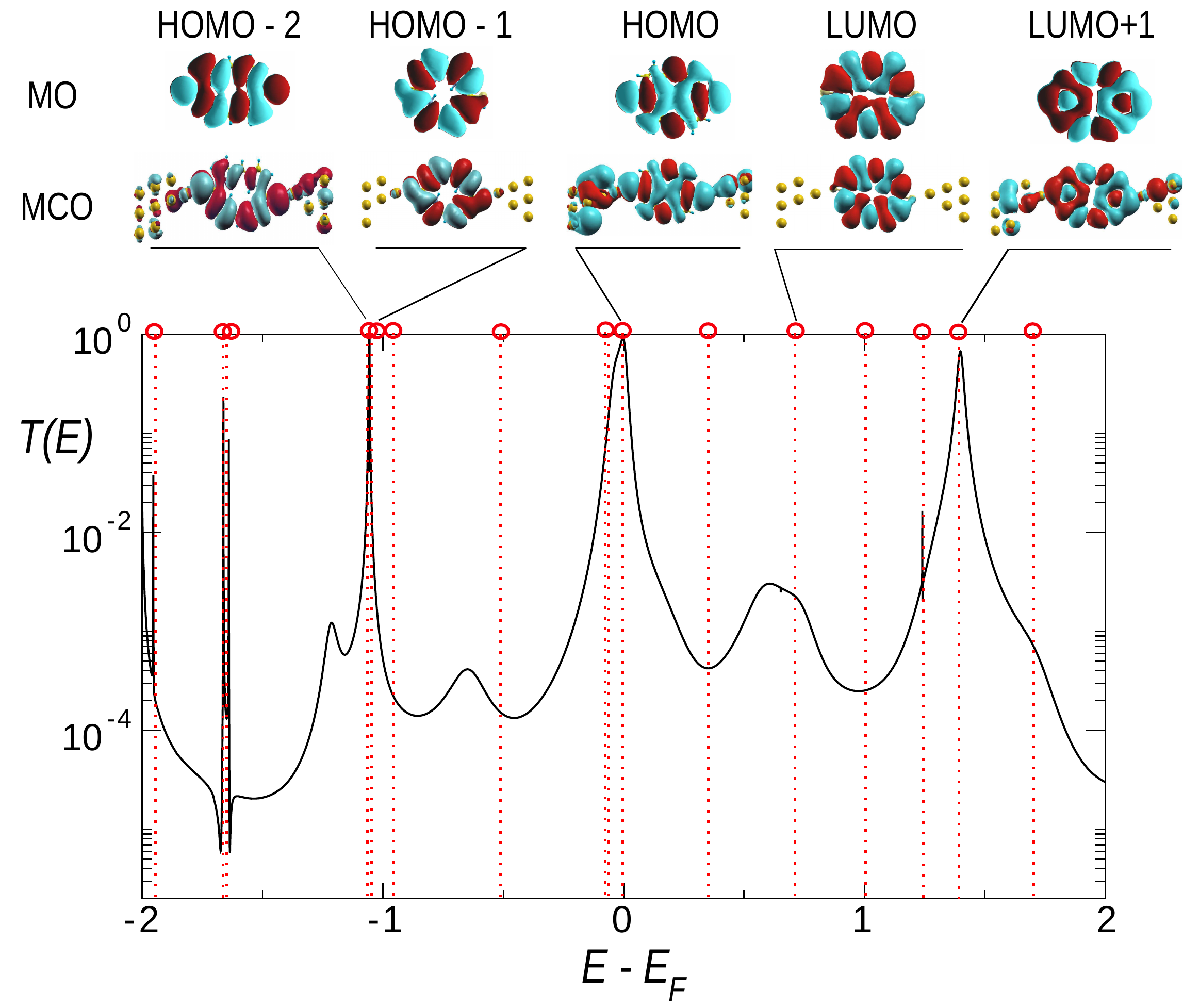}
    	\caption{The zero bias transmission function in normalized logarithmic scale for $a_1-a_2$ contact geometry. The red dashed lines and red circles indicate the MPSH energy levels, scaled with respect to the electrode Fermi energy. The wave function plots of selected MOs and MCOs are given at the top.} 
    \end{figure}

    For further understanding of the $I-V$ characteristics, we calculate the bias dependent transmission functions for all possible contact geometries. We present the transmission function under varying bias values for the $a_1-a_2$ and $a_1-a_4$ cases in Figure 4a and b, respectively, and choose to present the results for rest of the systems in SI. Note that, for comparison of transmission among all the systems, we present the absolute transmission values in log-scale without any normalization. The shaded regions in the figures depict the bias window of $[-V/2,V/2]$, i.e., the integration limits for the current calculation. As can be seen in Figure 4a, the transmission peak at Fermi energy for $+0.1V$ entirely falls within this integration limit, leading to a peak in current response at that bias. At $-0.1V$, this transmission peak slightly shifts below the Fermi energy and consequently the current response slightly diminished as compared to the previous case. Further increase in bias values pushes the transmission peak further away from the Fermi energy and outside of the bias window, causing a fall of current response. In case of $+0.8V$, the transmission peak enters the bias window, which explains the increase in the current around that bias value, as has been observed in Figure 2a. A close look in all the plots reveal that the transmission values under the forward biases is larger than that under the reverse biases. This results into asymmetric current response and possible rectification behavior that we have observed. 

    To investigate the low current response of the $a_1-a_4$ system, we present its bias dependent transmission functions in Figure 4b. As can be seen, the transmission values in the forward bias direction within the bias window is substantially higher than that in the reverse bias. This explains the negligible current in the reverse bias. In the forward bias too, the overall transmission peak height in the bias window is considerably lower than that in the other systems, leading to comparatively weak current response of this system. 

    As mentioned before, the electronic conduction through the system mainly happens through the HOMO for all contact geometries. In order to understand the varying current responses, we present the MCOs corresponding to the HOMO states for all the contact geometries in Figure 4c. The maximum wave-function overlap over the whole scattering region leads to maximum current response in case of $a_1-a_2$ contact geometry. The reduction in wave-function overlap causes the decreased current response in $a_1-a_3$, $a_1-a_5$ and $a_1-a_6$ systems. In case of $a_1-a_4$ system, the MCO shows negligible delocalization over the right electrode, unlike the other contact geometries. This leads to the decreased transmission near the Fermi energy and corresponding negligible current.

    Furthermore, it can be noticed from the Figure 4c that for all systems, the corresponding wave-function overlap of the molecular region with right electrode is substantially lower than that with the left electrode. This accounts for the asymmetric current responses in the forward and reverse bias values in all the contact geometries. Among all the systems that we investigate here, the $a_1-a_2$ contact geometry ensures symmetric attachments of the thiol anchoring groups in either sides of the symmetric porphine molecule. Such symmetric system is expected to exhibit symmetric $I-V$ characteristics under the forward and reverse biases, when the system is placed between two same electrodes in either sides. The asymmetric transport behavior has been reported to arise from the asymmetry in molecular geometry, either arising from the asymmetric structural backbone\cite{ding2014single} or due to the attachment of donor and acceptor groups to the molecule\cite{van2015molecular,pan2011current,thong2017homo}. Such asymmetric transport behavior can also arise from asymmetric attachment of the molecule to the electrodes via anchoring groups\cite{zhao2010molecular} or due to use of two structurally or chemically different electrodes in either sides\cite{jiang2016effects,kaur2016effect}. In the present study, the usage of tip electrodes in either sides of the molecule creates structural asymmetry in left and right sides, as can be seen in Figure 1b. This might lead to different self-energies for two electrodes, arising from asymmetric coupling of the system with the left and right electrodes, leading to asymmetric $I-V$ characteristics. To investigate further about this, we model our system within tight-binding theoretical framework and impose equal self-energies in either sides of the molecule.

    \begin{figure}[h]\centering		
    	\includegraphics[scale=0.27]{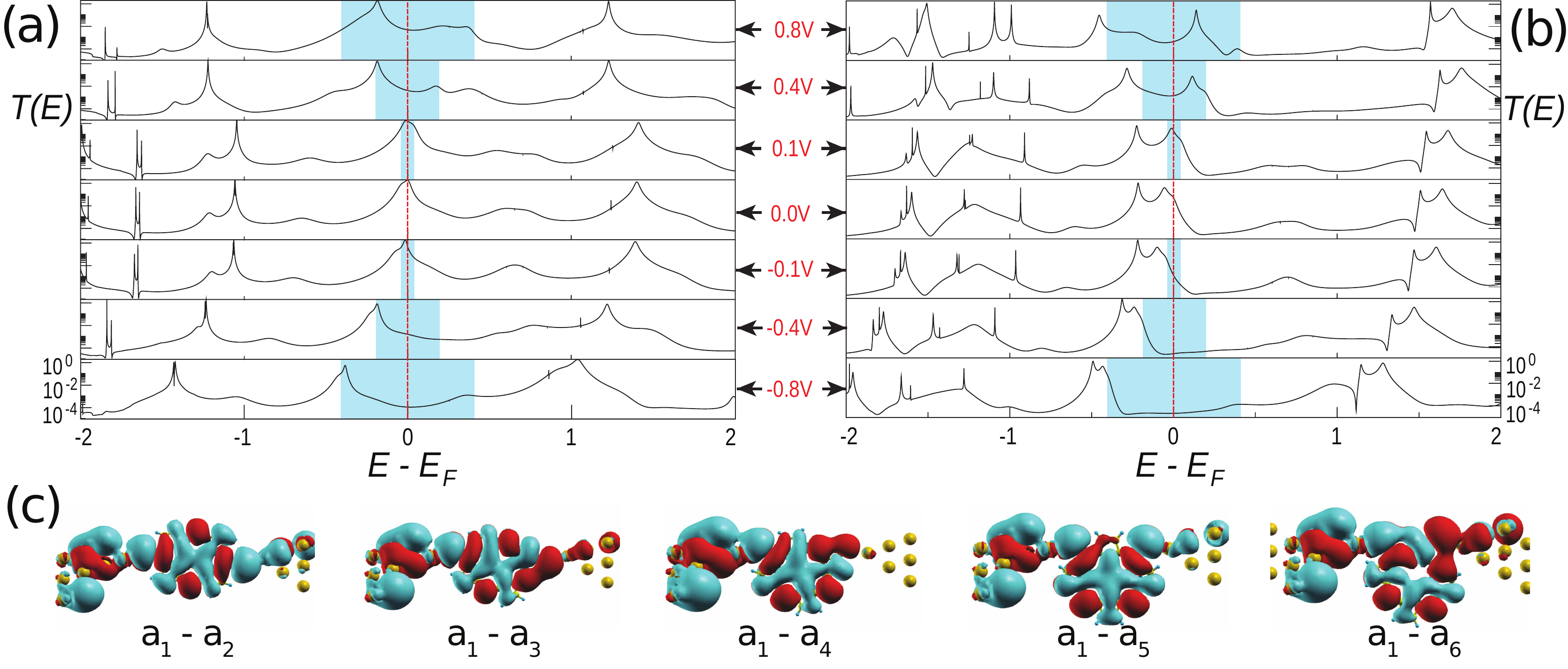}
    	\caption{The non-zero bias transmission functions (in logarithmic scale) of (a) $a_1-a_2$ and (b) $a_1-a_4$ contact geometries. The bias values are mentioned beside the corresponding transmission plot panel. The red dashed line indicates the Fermi energy and the shaded boxes depict the bias window. (c) The MCOs corresponding to the HOMO states at zero-bias for different contact geometries.}   	
	\end{figure}

    The $\pi$-electron tight-binding hamiltonian for the molecular system is expressed as:
	\begin{equation}
	    H_{mol} = \sum_{i,\sigma}\epsilon_i c_{i,\sigma}^\dagger c_{i,\sigma} + \sum_{<ij>,\sigma} t(c_{i,\sigma}^\dagger c_{j,\sigma}+c_{j,\sigma}^\dagger c_{i,\sigma}) - \sum_{i,\sigma}V_i c_{i,\sigma}^\dagger c_{i,\sigma}
	\end{equation}
	
    \noindent where $c_{i,\sigma}^\dagger$ ($c_{i,\sigma}$) creates (annihilates) an electron with spin $\sigma$ at $i$-th site. The $\epsilon_{i}$ and $t$ are the onsite potential and nearest neighbor hopping integral. Note that, for tight-binding calculations, we consider only the porphine molecule without the thiol anchoring groups, since the effect of the same can be included later in terms of the self-energy. The term $V_i$ defines the site potential, arising from the applied external bias that is modeled as a ramp and can be written as\cite{treboux1998asymmetric}:
	\begin{equation}
	V_i=-\frac{V_L r_{iL} + V_R r_{iR}}{r_{iL}+r_{iR}}
	\end{equation}
	
    \noindent where $V_{L/R}$ is the potential at the corresponding contact-site, attached to $L/R$ electrode and $r_{i(L/R)}$ is the vectorial distance between $i^{th}$ site and the respective electrode. The value of $V_{L/R}$ is considered as $\pm V/2$, where $V$ is the potential difference between left and right electrodes. We set the $t$ as the unit of energy for our calculations. We consider the onsite energy of carbon atoms, $\epsilon_C=0$, whereas the onsite energy of nitrogen atom is taken as $\epsilon_N=-0.5$, owing to its higher electronegativity as compared to the carbon. We use identical self-energies ($\Sigma_{L/R}$)  of $0.05$ on either sides of the porphine molecule to account for the electrodes including the effect of the anchoring groups. Thereafter, we integrate the tight-binding hamiltonian with the NEGF formalism to calculate the transmission and the current as a function of the applied bias.
	
	We plot the $I-V$ characteristics of porphine molecule for varying contact geometries in Figure 5a. As can be seen, the symmetric contact geometry $a_1-a_2$ now shows symmetric $I-V$ plot in $+ve$ and $-ve$ biases. This clearly proves that the structural asymmetry of the electrodes in either sides, as considered within DFT calculations is the sole reason for asymmetric $I-V$ plot for $a_1-a_2$. Interestingly, the $a_1-a_5$ also shows symmetric $I-V$ plot due to both electrode contacts on methine bridge, similar to the $a_1-a_2$. However, the asymmetric contact geometries in other cases results in slight asymmetry in the $I-V$ characteristics. Note that, the device geometry in Figure 1b that is considered for the DFT calculations, is more close to the experimental setup. Therefore, it can be inferred that, the current responses towards the $+ve$ and $-ve$ biases for any molecule can be asymmetric under experimental measurements, irrespective of the inherent symmetry of the molecule.
	
	In Figure 5b, we present the normalized transmission functions for all the contact geometries at zero bias. All the contact geometries exhibit large transmission at and near the Fermi energy, except the $a_1-a_4$ system which shows wide antiresonance at and near the Fermi energy. This is a signature of either destructive interferences or negligible constructive interferences among the MOs that take part in the transmission. As a result, the current response of $a_1-a_4$ system becomes lowest among all. This is consistent with our observations from DFT calculations as well. Note that, this prominent antiresonance feature is not visible in the DFT calculations, due to the consideration of $\sigma$ electrons therein, as have been reported earlier\cite{ke2008,solomon2009electron}. The maximum transmission of $a_1-a_2$ system makes its current response highest. Note that, the other contact geometries also exhibit the antiresonance away from the Fermi energy and comparatively less wider than that of the $a_1-a_4$ system. This feature along with their overall transmissions at and near the Fermi energy decide their respective current responses.
	
	To gain further insight about the interferences that result into such antiresonance behavior, we further explore the quantum interference (QI) effect as prescribed by Gunasekaran et.al.\cite{gunasekaran2020}. The Green's function is typically described in atomic orbital (AO) basis. A basis transformation of the Green's function from AO to MO basis is necessary to extract the QI information. This involves a transformation matrix $P$, the columns of which are comprised of the eigenvectors of $G$ such that $P^{-1}GP$ is diagonal. In this MO basis, the transmission function $T$ can be written as,
	
	\begin{equation}
	T=\sum_{ij}Q_{ij}
	\end{equation}
	
	Here, $Q_{ij}$ is $(i,j)$-th element of the $Q$-matrix which is defined as,
	
	\begin{equation}
	Q(E)=(P^\dagger \Gamma_L G P)\circ(P^{-1} \Gamma_R G^\dagger P^{-1 \dagger})^T
	\end{equation}
    \noindent where $``\circ"$ denotes the entry-wise (Schur) product, and $(*)^T$ is matrix transpose. 
	
	As can be seen from the above equations, the transmission function and the $Q$-matrix are equivalent to each other. For visual representation of the QI effect, we plot the color-map of the zero bias $Q$-matrix at the Fermi energy in Figure 5c. We choose to plot only few frontier orbitals among which the HOMO is doubly degenerate, marked with subscripts $``a"$ and $``b"$. The diagonal elements of the $Q$-matrix depict the individual contribution of a MO towards the transmission, whereas the off-diagonal elements represent the interferences among the MOs. Note that, the diagonal has been taken from the bottom-left to top-right corner in the color-maps, as obtained from the Mathematica package\cite{Mathematica}.

  	\begin{figure}		
    	\includegraphics[scale=0.25]{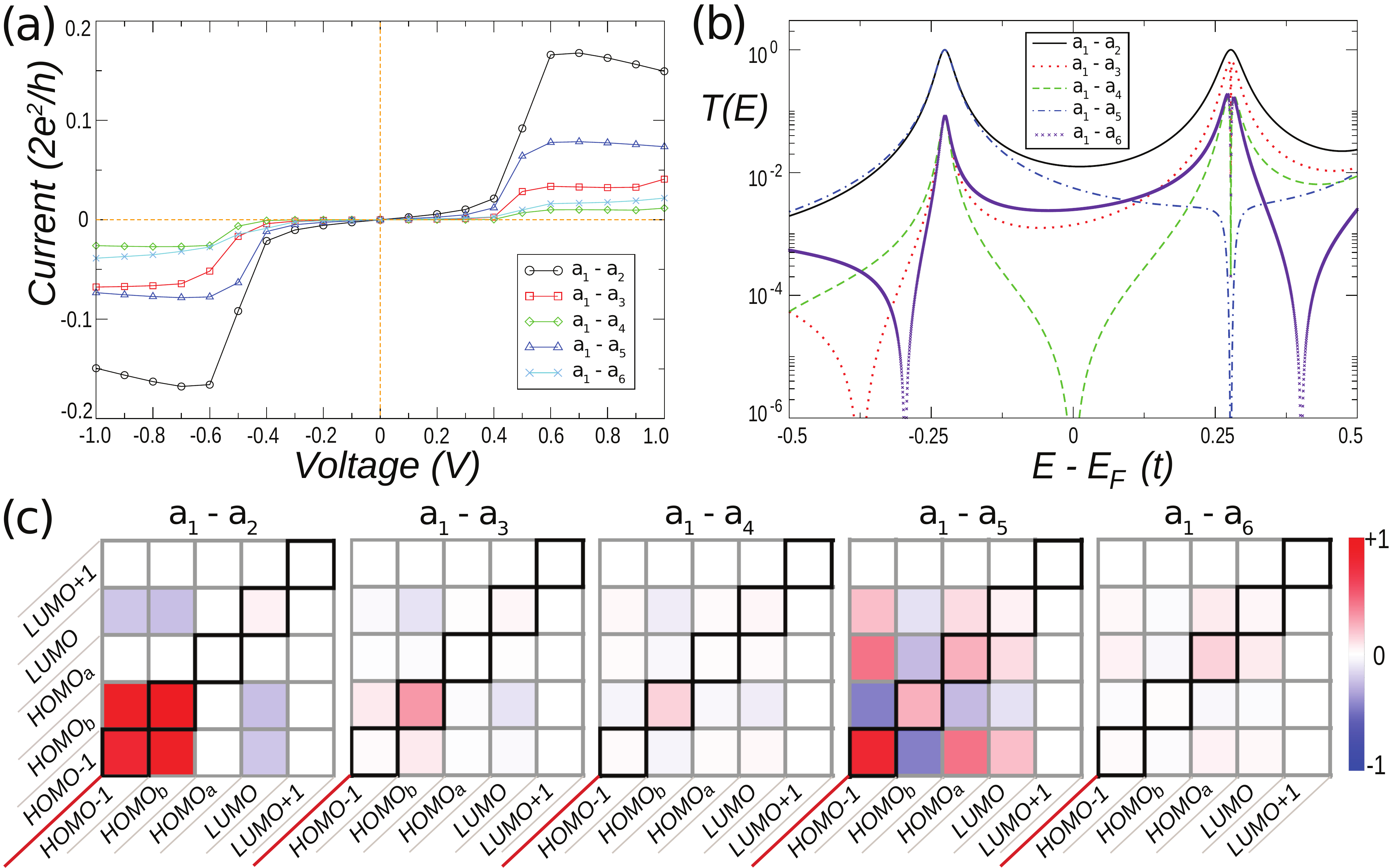}
    	\caption{(a) The current-voltage characteristics and (b) the zero bias transmission functions (in normalized logarithmic scale) of all the contact geometries, as obtained from tight binding and NEGF calculations. (c) Color-map representation of the normalized Q-matrices of corresponding systems for few frontier MOs. The red (blue) color indicates constructive (destructive) interference.}
	\end{figure}
	
	The largest constructive interference between the HOMO-1 and one of the HOMOs of $a_1-a_2$ system results in highest transmission at Fermi energy. The $a_1-a_5$ system also exhibits constructive interference among few of the frontier MOs, albeit less than the previous one. Moreover, this system shows considerable destructive interference between the HOMO-1 and HOMO$_b$ as well, resulting in overall decrease in the transmission as compared to the $a_1-a_2$ system. The magnitude of the constructive interferences in rest of the three systems are comparatively smaller than the previous two, thereby the transmission of these three systems drop. In case of $a_1-a_4$ system, the HOMO$_b$ contributes towards the transmission although with one order of magnitude less than that of the $a_1-a_2$ system. However, the destructive interferences between the HOMO-1, HOMO and LUMO levels result in antiresonance at Fermi energy for this system. Note that, the unified color scale for all the $Q$-matrix color-maps does not portray these destructive interferences clearly in Figure 5c. That is why, we provide the same with individual color scales in Figure S6. These $Q$-matrix representations clearly show the MO contributions and the QI effect among the MOs that govern the transmission behavior and the overall current response for different contact geometries of porphine.

   Note that, all the above discussions and inferences are consistent for the $``b"$-type contact geometries as well (see SI). Based on our results we can propose the sensor device fabrication using biomolecules with porphine building blocks. As we know, the chlorophyll and haemoglobin are two major constituents that control important physiological activities of plants and animals, respectively. Any irregularity in such biomolecules leads to disruption of such activities. Such irregularities are expected to result in a unique current response and hence, can be detected in terms of the $I-V$ characteristics through the sensor device based on the proposed setup herein.

\section{Conclusion}

We investigate the electronic transport through the porphine molecule with varying contact geometries using the NEGF formalism coupled with both DFT and TB computational framework. To model the experimental device geometry, we consider a tip electrode that mimics the mechanical break junction. This allows electrode attachment to the specific contact sites of the molecule with atomic precision. With such device configuration, we observe varying transmission and current responses corresponding to different contact geometries. Specifically, $a_1-a_2$ system shows highest current, whereas the current response from the $a_1-a_4$ system is negligible. Such varying responses originate from the different molecular orbital contributions and the delocalization of the wave-functions over the device region that are further established based on MPSH calculations. Interestingly, we observe asymmetric $I-V$ characteristics under the $+ve$ and $-ve$ biases for all the contact geometries, irrespective of the symmetry of the molecule in between two gold electrodes. We attribute this to the structural asymmetry of the tip contacts in either sides of the molecule. We establish this phenomena within tight-binding calculations by considering identical self-energies to model the left and right electrodes. Further investigations of quantum interference effect among the frontier orbitals based on $Q$-matrix approach reveal the origin of varying transmission features and the consequent $I-V$ characteristics for all the contact geometries. Note that, the observation of asymmetric $I-V$ characteristics can be generalized to any molecular system, owing to the structural difference between the source-drain electrodes under experimental setup. This will provide microscopic understanding of the current responses and will drive precise device engineering. Furthermore, owing to the abundance and importance of the porphine molecular system in numerous biomolecules and biophysical activities, our study can provide guidance for the realization of sensor devices.

\begin{acknowledgement}
The authors thank IISER Tirupati for Intramural Funding and SERB, Department of Science and Technology (DST), Govt. of India for research grant CRG/2021/001731.
\end{acknowledgement}

\begin{suppinfo}
	Supporting Information (SI) contains the systems considered for calculation and their method of preparation for DFT calculations, $I-V$ characteristics and conductance for $a$ and $b$ type contact geometries, non-zero bias transmission functions as obtained from DFT calculations and TB calculations with the color maps of $Q$-matrix for respective systems. 


\end{suppinfo}

\bibliography{refer}
\clearpage


\begin{center}
	\textbf{  \Huge {SUPPORTING INFORMATION}}
\end{center}

   \begin{figure}[h]\centering
    	\includegraphics[width=0.85\textwidth]{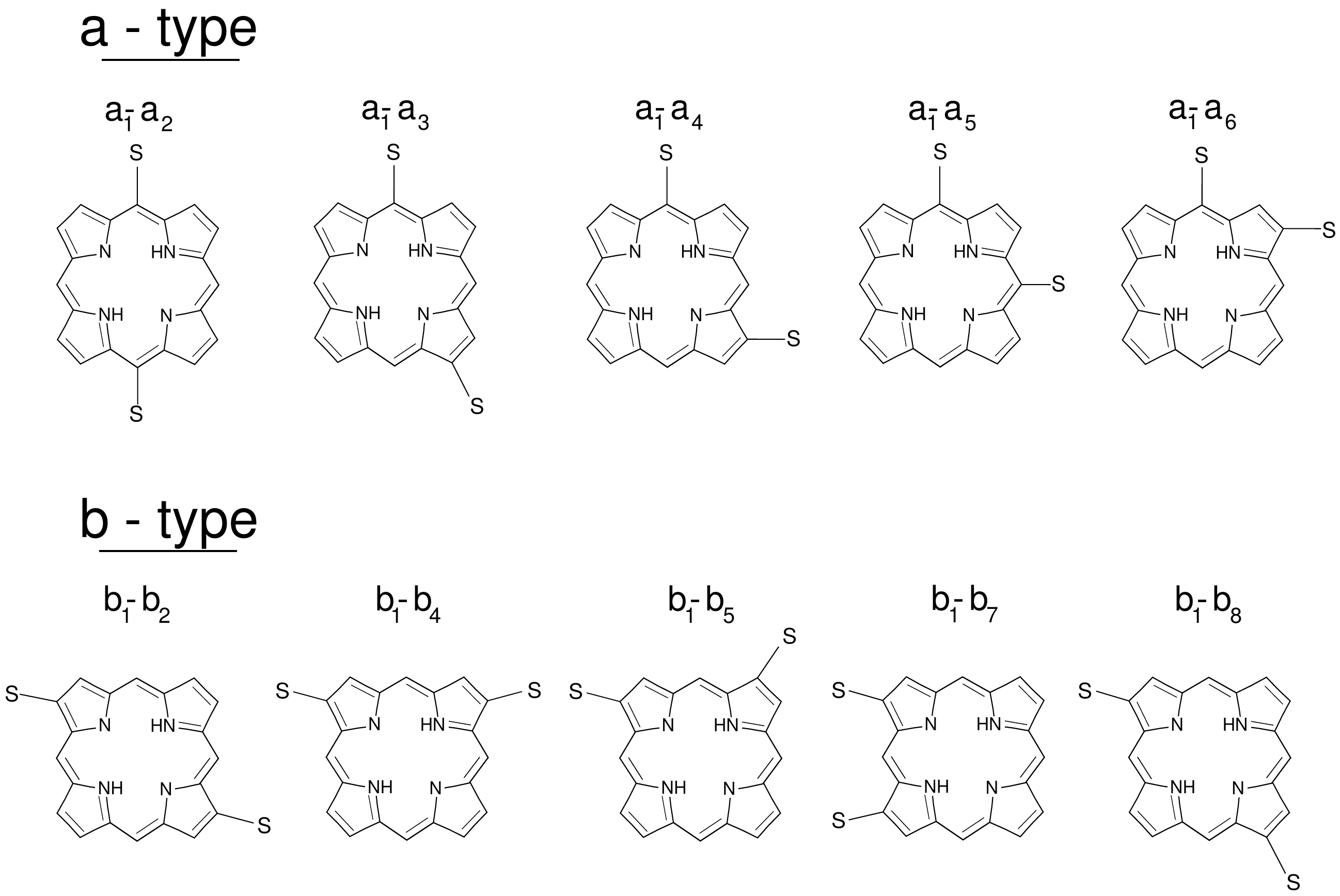}
	    \caption*{\textbf{Figure S1} : Ten different contact geometries of "$a$" and "$b$" types. The tip electrodes are attached to the thiol anchoring groups. Type "$a$" contact geometries are having at least one electrode attached to the methine bridge, whereas, the type "$b$" contact geometries are having at least one electrode attached to the pyrrole rings. Note that, $a_1-a_3$ and $a_1-a_6$ are identical to $b_1-b_3$ and $b_1-b_6$, respectively. That is why, only the "$a$" type contact geometries are chosen.}
    \end{figure}

\section{Computational details}   
	The structure of the porphine molecule is first  optimized using density functional theoretical framework, as implemented in the SIESTA-3.2 package\cite{soler2002siesta}. Then the thiol groups are attached at different sites on the molecule to obtain the geometries in Figure S1, followed by further geometrical optimization of the full systems. In the third step, each of the optimised geometries are placed in between a pair of pyramidal gold tips made of 4 $Au$ atoms, which are optimised again to obtain the relaxed $S-Au$ distances with the Au-atoms being frozen during the runs. This gives us the extended molecule comprising of porphine molecule, thiol anchoring group and $Au$ tip electrodes.

	In the next step, the extended molecule is placed between 1-2 layers of $Au$-atoms, thus makng the scattering region. Then the scattering region is placed between a pair of identical $Au(111)$ bulk-electrodes. This whole system is now considered to be the computational cell used for the transport calculations. The gold tips are attached to the bulk gold electrodes in a manner so that they appear like sculptured from the bulk electrodes. For the transport calculations, we first calculate the transmission functions using TranSIESTA\cite{stokbro2003transiesta} follwed by the current calculations using TbTrans. Both the modules are included in the SIESTA-3.2 package. 

	Structural visualisations of the systems are done using GaussView6. The molecular orbital visualisations are done with the XCrySDen.

	\begin{figure}[h]\centering
		\includegraphics[width=1.0\textwidth]{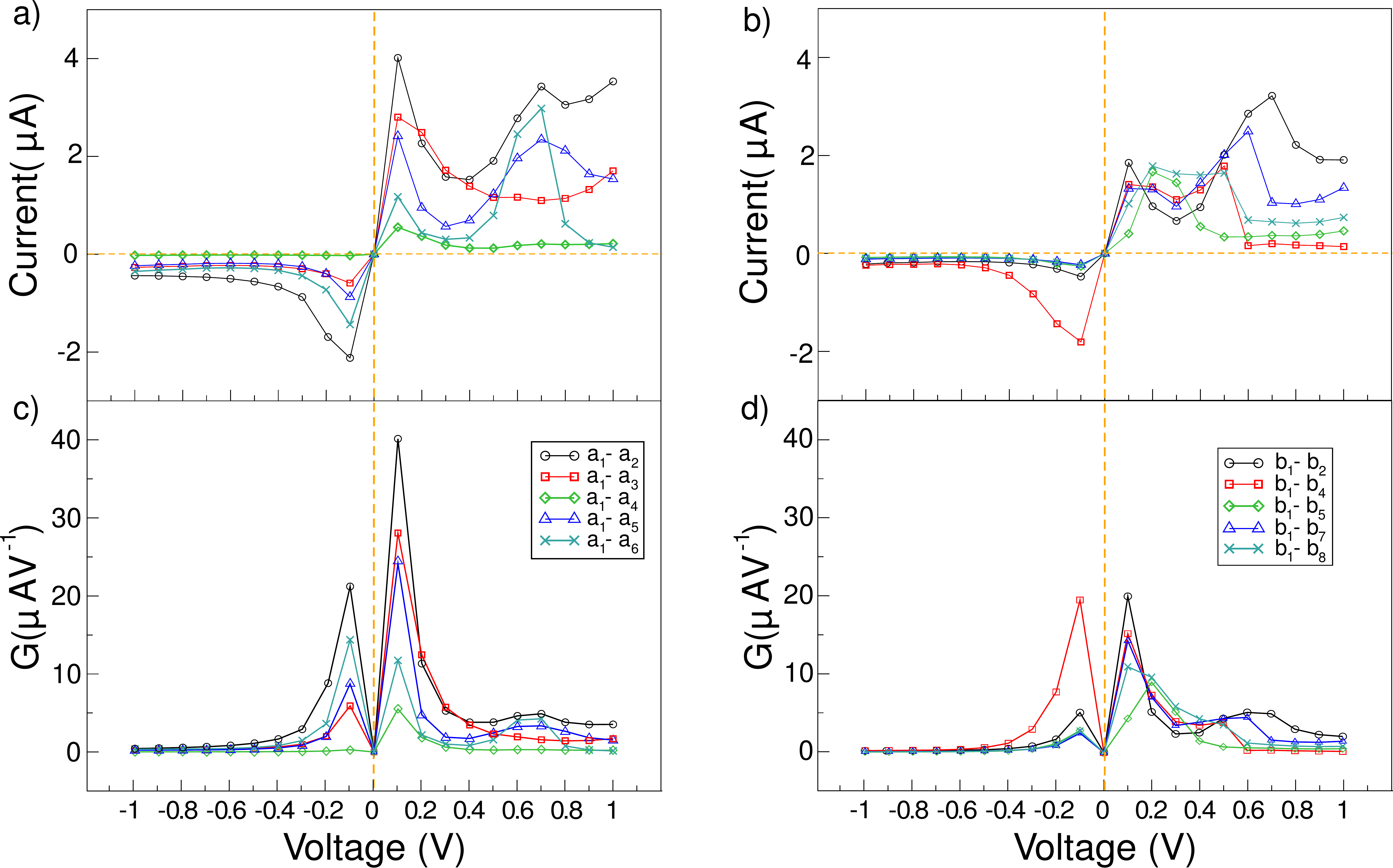}
		\caption*{\textbf{Figure S2} : The current-voltage characteristics of (a) "$a$" type and (b) "$b$" type contact geometries. Their conductance plots are given in (c) and (d), respectively.}
	\end{figure}

	\begin{figure}[h]\centering
		\includegraphics[width=0.8\textwidth]{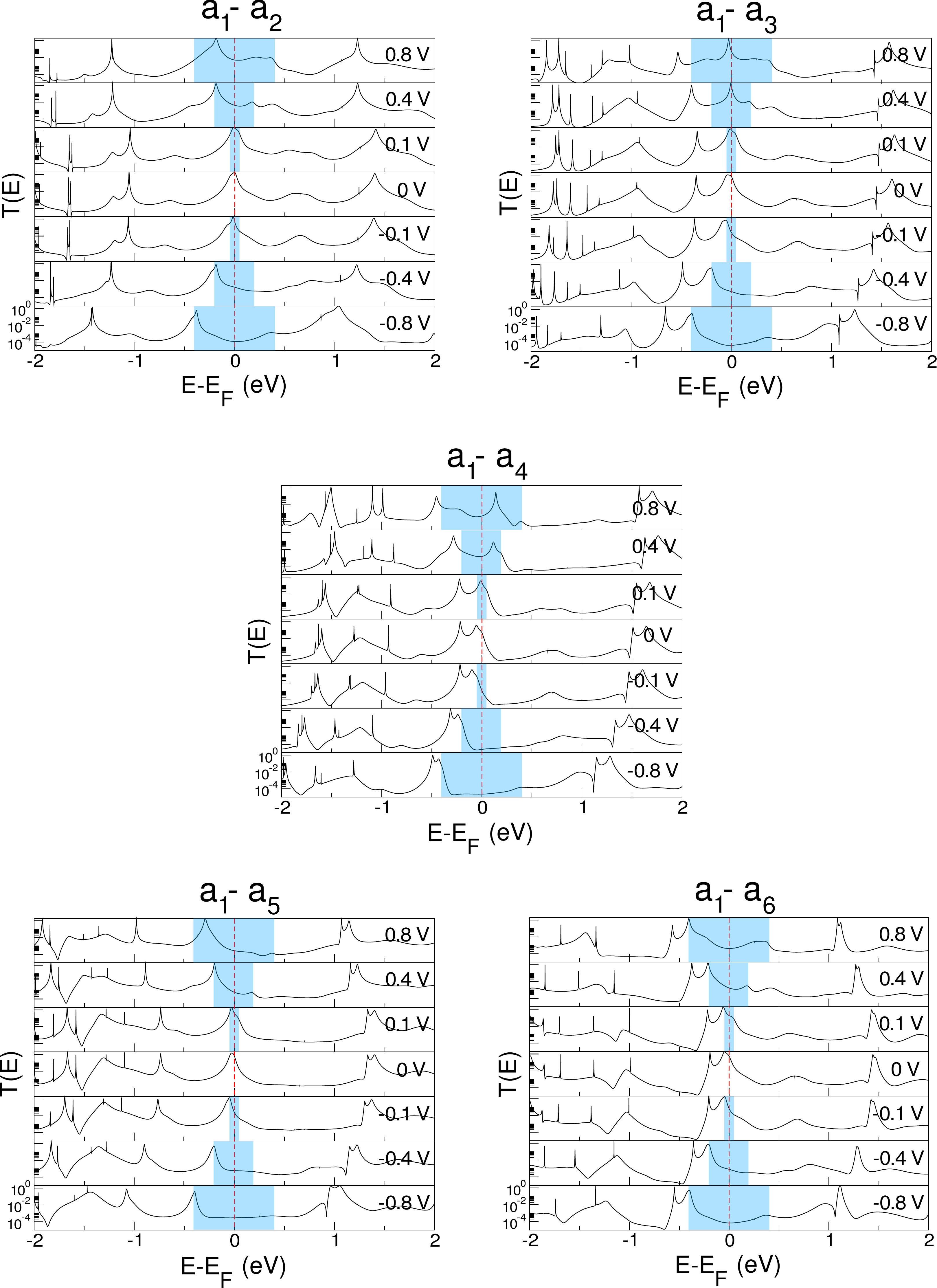}
		\caption*{\textbf{Figure S3} : The non-zero bias transmission functions obtained from DFT calculations for all the "$a$" type contact geometries. The bias values are mentioned in corresponding panels. The shaded regions indicate the bias window.}
	\end{figure}

	\begin{figure}[h]\centering
		\includegraphics[width=0.8\textwidth]{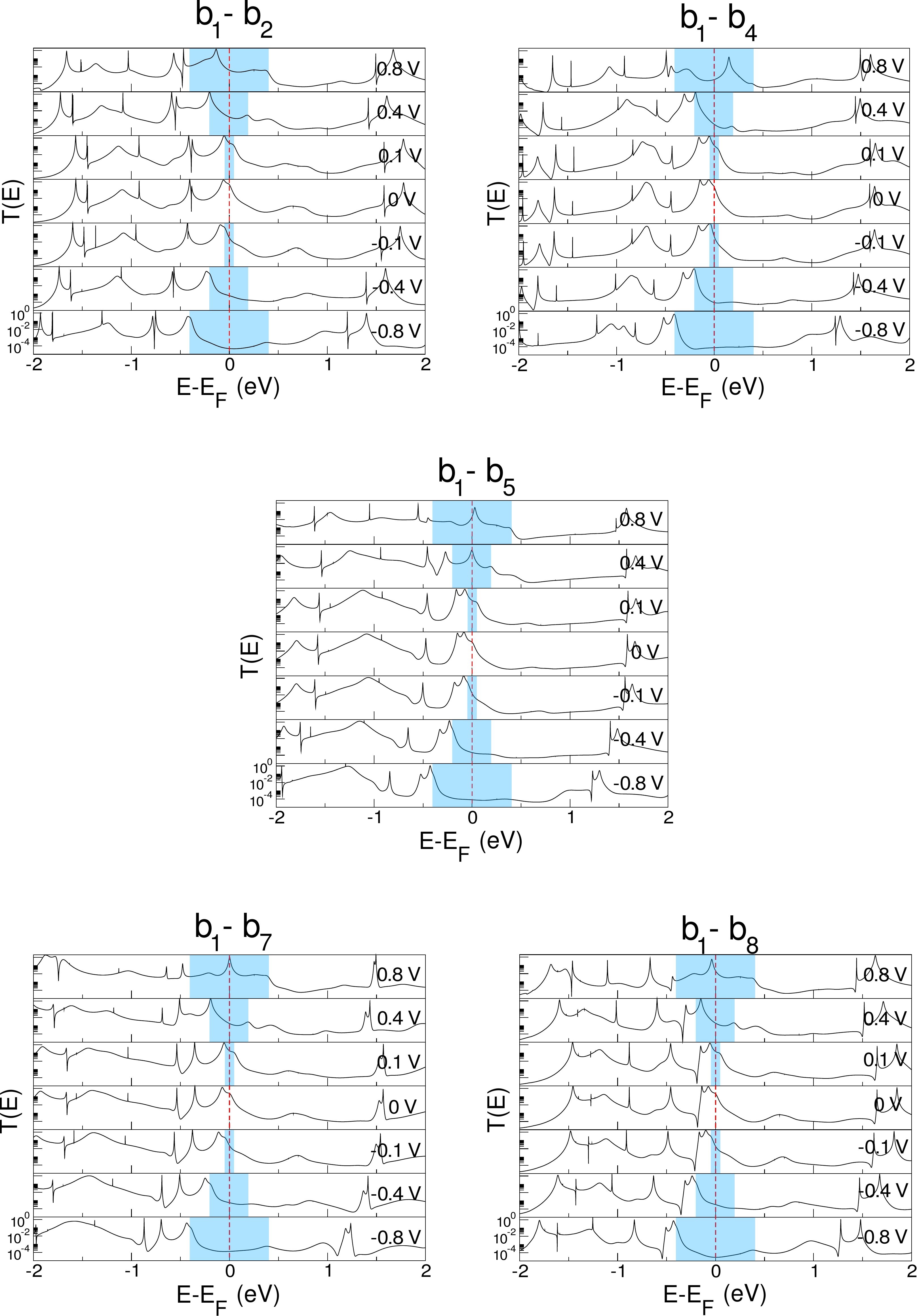}
		\caption*{\textbf{Figure S4} : The non-zero bias transmission functions obtained from DFT calculations for all the "$b$" type contact geometries. The bias values are mentioned in corresponding panels. The shaded regions indicate the bias window.}
	\end{figure}

	\begin{figure}[h]\centering
		\includegraphics[width=1.0\textwidth]{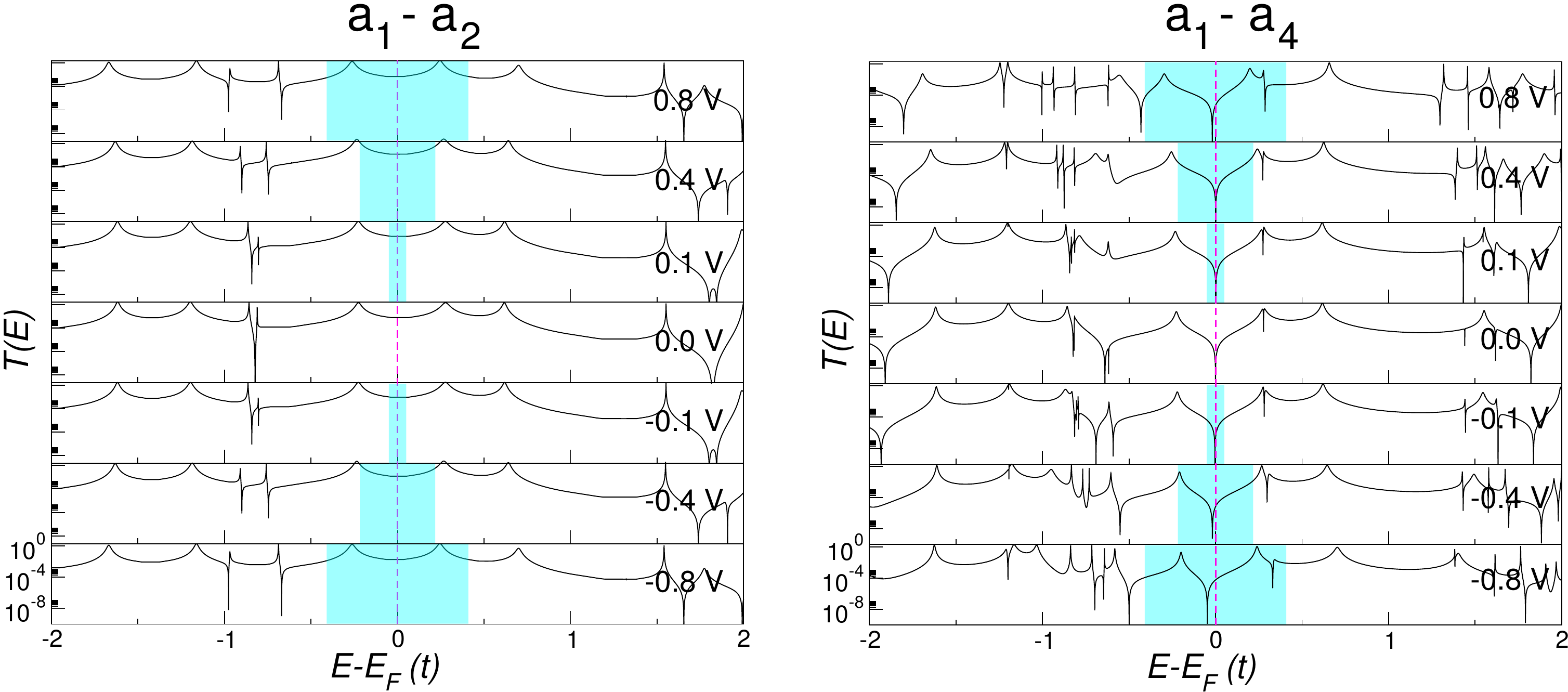}
		\caption*{\textbf{Figure S5} : The non-zero bias transmission functions obtained from tight-binding calculations for $a_1-a_2$ and $a_1-a_4$ systems. The bias values are mentioned in corresponding panels. The shaded regions indicate bias window.}
	\end{figure}

	The $Q$-matrices are color-mapped for visual understanding of the interferences among the frontier MOs and using them to interpret the transmission functions. These plots are prepared using the Mathematica package\cite{Mathematica}. Red indicates constructive interference and blue indicates destructive interference. Note the color scale in Figure S6 and S8 are different for each system. The constructive interference for $a_1-a_2$ is an order of magnitude higher than that of $a_1-a_4$, but they appear to be of the same color-shade. The plots in Figure S6 and S8 emphasize the fact that both constructive and destructive interferences among MOs are possible. For one-to-one comparison among the systems, normalization of Q-matrix values are necessary which leads to very faint colors as has been represented in Figure 5c of the main manuscript. The net effect of the interferences in a system gives a direct insight into the features of the transmission function at that specific energy point and bias value. The antiresonance observed for $a_1-a_4$ at Fermi energy (presented in Figure 5b of the main manuscript) is due to the significant destructive interference between HOMO$_b$ and (HOMO-1), HOMO$_a$ and LUMO, which cancels out the weak constructive interferences among the MOs. All discussions for "$a$" type contact geometries equally hold for the "$b$" type systems.

	\begin{figure}[h]\centering
		\includegraphics[width=1.0\textwidth]{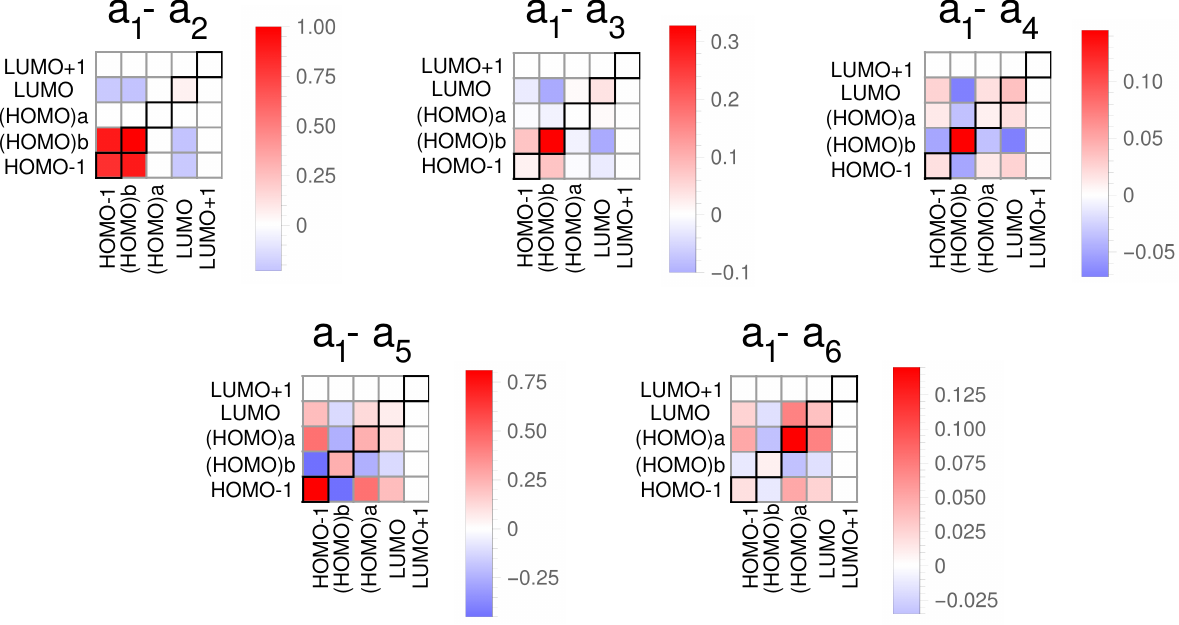}
		\caption*{\textbf{Figure S6} : Color maps depicting the Q-matrices for the "$a$" type contact geometries at the Fermi energy for zero-bias. Note that the colour scale is normalized for $Q_{max}$ for individual system.}
	\end{figure}

	\begin{figure}[h]\centering		
		\includegraphics[width=1.0\textwidth]{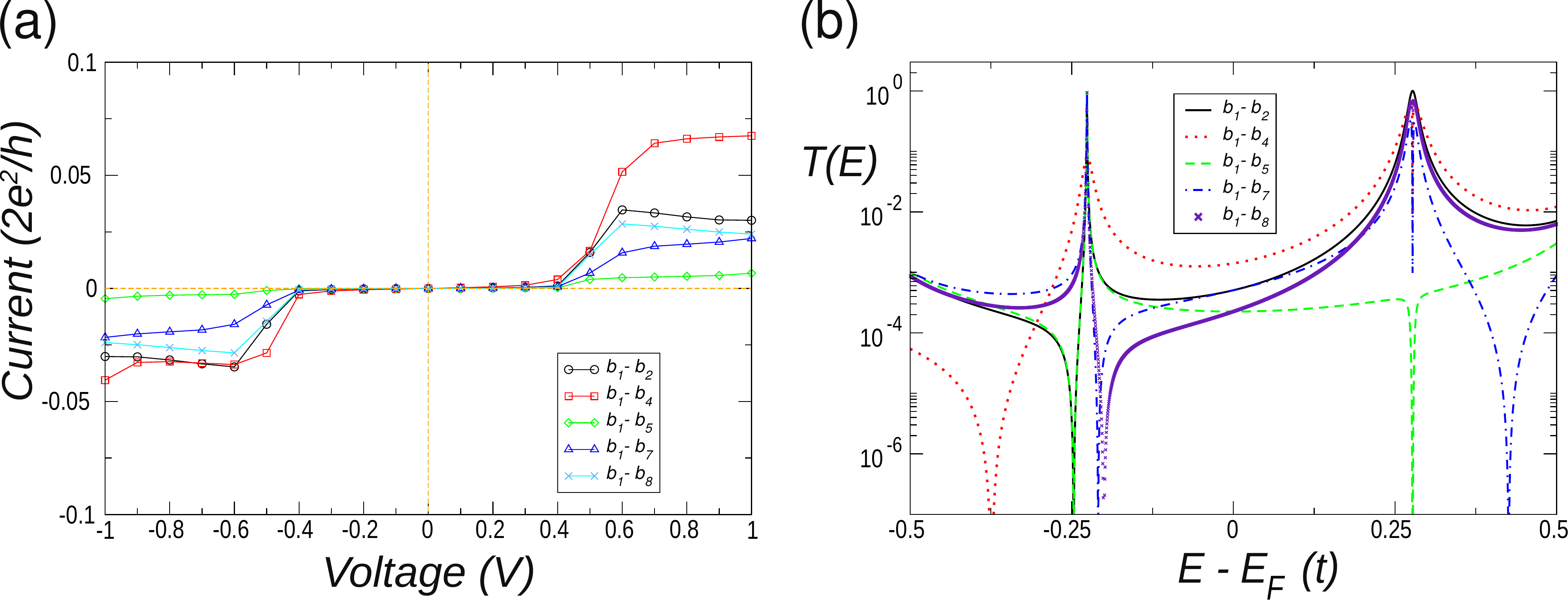}
		\caption*{\textbf{Figure S7} : (a) The current-voltage characteristics and (b) the zero bias transmission functions (in normalized logarithmic scale) of all the "$b$" type contact geometries, as obtained from tight binding and NEGF calculations.}
	\end{figure}

	\begin{figure}[h!]\centering
		\includegraphics[width=1.0\textwidth]{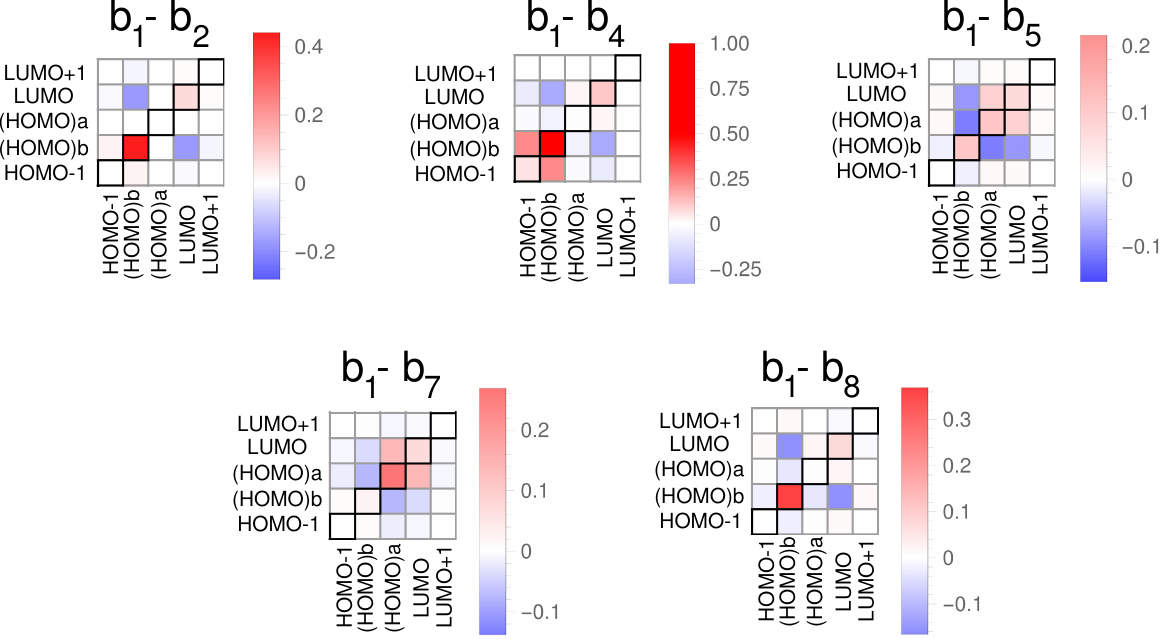}
		\caption*{\textbf{Figure S8} : Color maps depicting the Q-matrices for the "$b$" type contact geometries at the Fermi energy for zero-bias. Note that the colour scale is normalized for $Q_{max}$ for individual system.}
	\end{figure}

\end{document}